\documentstyle[proceedings,numreferences]{crckapb}
\input psfig
\pssilent
\begin{opening}
\title{Colliding black holes: analytic insights}
\author{Jorge Pullin}
\institute{
Center for Gravitational Physics and Geometry\\
Department of Physics, Pennsylvania State University,\\
104 Davey Lab, University Park, PA 16802}
\end{opening}
\begin{document}
\begin{abstract}
We summarize the state of the art of the ``close approximation'' to 
black hole collisions. We discuss results to first and second order in
perturbation theory for head-on collisions of momentarily-stationary
and non-stationary black holes and discuss the near-future prospect of
non-axisymmetric collisions.
\end{abstract}
\vspace{-10.5cm} 
\begin{flushright}
\baselineskip=15pt
CGPG-98/2-4  \\
gr-qc/9803005\\
\end{flushright}
\vspace{9cm}

\section{Introduction}

When two black holes collide, one can distinguish three regimes in the
collision with distinct behaviors. Initially, when the black holes are
far apart, one can think of them as individual black holes, each with
their own horizon. This regime can be well approximated by
post-Newtonian analysis \cite{PN}. As the black holes approach, they
start exerting more significant influence on each other and their
structure becomes distorted. Some approximations have also been
suggested for this regime \cite{Suheat}. At some point however, the
influence is too intense for approximations to work and one enters a
fully nonlinear regime, generally perceived as only treatable by full
numerical simulations (from here on we will use the name ``full
numerical'' to mean numerical simulations of the Einstein equations
without approximations). As is well known, a great numerical effort is
currently underway to try to model this regime, a summary being
present in this same volume in the talk by Ed Seidel. Towards the end
of the collision, the two black holes will not really be two black
holes any longer, but will merge into a larger, highly distorted
single black hole. This latter part of the collision could be treated
using perturbation theory of a single black hole. This has been known
for quite some time. Four years ago, Richard Price and I set out to
explore this regime of the collision \cite{PrPu}. One can
alternatively view what we were trying to do as applicable to black
hole collisions for holes that start very close to each other. So
close that they are surrounded by a common horizon. Technically, they
are therefore a single hole. However, we expected that if we took the
same families of initial data that full numerical simulations were
using, and extrapolated them to the case of close holes, our technique
would yield results against which to benchmark the full numerical
simulations. This would make the results useful. What has been found
is that the approximation seems to work in a somewhat broader range of
parameters than was initially expected, allowing one to make
predictions for separations that are, at least in the head-on
collision, of the order of $L/m\sim 7$. At the same time, the advances
obtained in full numerical simulations have brought the field to a new
level of maturity and it is now recognized that there are several
important problems ahead before a full numerical in-spiral collision
will reliably be available for a large range of parameters. This makes
the close limit studies all the more valuable. For some time, they
will be the only source of (albeit only approximate) waveforms for
certain types of black hole collisions, just when the interferometric
detectors are supposed to come online. Moreover, they will be
important benchmarks for full numerical simulations. This warrants
going into some depth in the analysis of the close limit approximation
for collisions of black holes. This talk will provide an overview of
current efforts along these lines.

The close-limit portion of collisions of black holes is dominated by a
behavior characterized by a single hole ``ringing away'' the
distortions produced by the collision. The type of waveform one gets,
therefore, is dominated by the quasinormal ringing of a black
hole. These types of waveforms are not the best to be detected with
interferometric detectors. Although they are remarkably intense, they
consists of short, rapidly dying trains of oscillations, which are not
easily pattern-matched out of the noise. Moreover, for ordinary black
holes, they tend to be at a frequency much higher than that of typical
interferometric detectors. Are they observable at all? Hughes and
Flanagan \cite{HuFl}, and more recently Creighton \cite{Cr} have made
a detailed study of all possible phases of the collision and the
conclusion seems to be that for a signal to noise of about 8, and for
collisions with a final Kerr parameter of $0.98$, one would need
collisions with a mass of $100$ solar masses for the final ringing to
dominate over the in-spiral. This makes direct observation of the
waveforms we will discuss rather unlikely, but of course, not
completely ruled out. The final angular momenta of collisions is still
an unknown quantity, and if it is lower, it would help observability
of final ringing. That will also be the case if black holes with a
mass of $50$ Solar masses are more common than expected. For a
discussion along these lines see \cite{Pramaldi}. The conservative
attitude today is therefore that the kind of waveforms we will discuss
are probably of not great interest from the observability point of
view. A conservative philosophy can therefore be that the value of the
work to be presented arises as potential benchmarks for numerical
simulations. We will see that other surprises appear in the way. It
can also teach us in its own right things about generic behaviors of
black hole collisions and the initial data to be used. In summary,
although this approach is unlikely to have the final word about black
hole collisions, it can help us gain several unexpected insights. This
will be the theme of this talk.

The organization of this paper is as follows. In the next section we
will discuss the initial calculations that opened this whole approach,
based on a test case for momentarily-stationary black holes. We will
then describe how to endow the formalism with ``error bars'' using
second order perturbation theory. In section IV we describe the
collision of boosted black holes and black holes with spin and in
section V we discuss the prospect for in-spiralling collisions and
other issues.

\section{A test case: the Misner problem}

\begin{figure}
\centerline{\psfig{figure=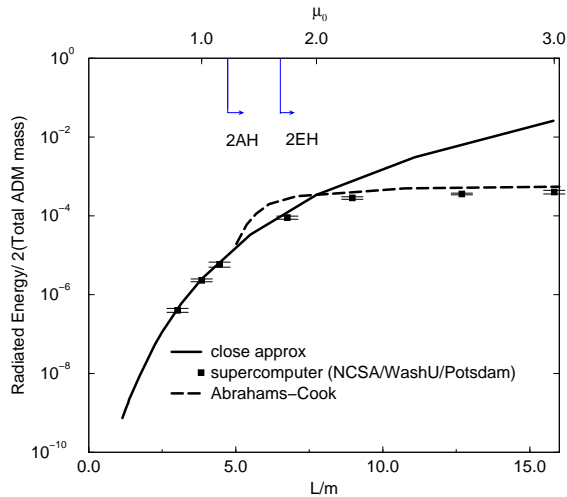,height=70mm}}
\caption{The energy released in a head-on collision of momentarily
stationary black holes. It is measured in fractions of the total ADM
mass of the initial slice and shown as a function of the initial
separation of the two holes. The latter is parameterized in two ways,
in the top it is given in terms of the $\mu_0$ parameter that appears
in the explicit form of the Misner geometry, in the bottom it is given
in terms of the separation of the throats in the flat conformal space
$L$ divided by one half of the ADM mass, as a measure of the ``mass of
each hole''. We indicate the regions in separation for which there are
two separate event and apparent horizons. We plot the results of the
close approximation, together with those of full numerical simulations
from the NCSA/Potsdam/WashU and the Abrahams-Cook approximation.}
\label{fig1}
\end{figure}

In the early 60's Misner \cite{Mi} presented an exact solution to the
initial value problem of general relativity. He noted that if one
imposes that the initial data be momentarily stationary, i.e., the
extrinsic curvature is zero, the constraint equations reduce to
requiring that the scalar curvature of the slice be zero. He then
constructed a conformally flat solution to that equation by obtaining
a conformal factor that solved the Laplace equation with boundary
conditions adapted to two holes connected by a throat. The
availability of this explicit analytic solution to the initial value
problem made it an ideal test case for studying the close
approximation: not only did we have exact initial data, but actually
the evolution of this problem in time has been pursued (given the
simplicity implied by axial symmetry) since the 60's by Smarr and his
group, and a detailed treatment has been presented in the last years
by the NCSA/Potsdam/WashU group \cite{NCSA}. So four years ago Richard
Price and I \cite{PrPu} looked at this in detail. Since the slice is
conformally flat, one readily finds coordinates in which, if the holes
are very close to each other, the exterior of the throats is isometric
to the Schwarzschild geometry in isotropic coordinates plus
perturbations. One then evolves the perturbations using the
traditional formalism for perturbations of a black hole. For this case
one finds that perturbations are pure $\ell=2$ in multipolar nature to
first order in the expansion parameter, and one can therefore evolve
them with the Zerilli equation. This is a linear differential equation
for a coordinate invariant combination of the metric perturbations
that has the structure of a Klein--Gordon equation in $1+1$ dimensions
with a potential. The potential represents the back-scattering of
perturbations by the curved geometry of the black hole. The equation
is written in terms of the radial $r*$ tortoise coordinate, which
covers only the exterior of the black hole, the horizon being at
$r*\rightarrow -\infty$. This does not allow the formalism to answer
straightforwardly questions about horizon dynamics, etc. The results of
evolving the perturbations determined by the initial data given by the
Misner geometry are summarized in figure \ref{fig1}.
In the figure, the continuous curve represents the energy computed
using the close approximation. It is worthwhile emphasizing that this
curve has {\em no free parameters}, that is, once given the initial
separation, the Misner data choose a unique evolution and radiated
energy. As can be seen the agreement with the numerical results for
this problem of the NCSA/Potsdam/WashU group \cite{NCSA} is quite good
up to a little after the two holes stop being separated by a common
event horizon. That is indicated by the two vertical lines at the top
of the plot. To the right of them, there are no common horizons of the
type indicated. Given the simplicity of the close limit computation,
the results are rather remarkable. Clearly this is a useful benchmark
for numerical relativity.

Also shown is the Abrahams and Cook approximation \cite{AC}. 
Theirs is an extension of the close approximation. Their rationale is
as follows: suppose you want to collide two black holes at $L/m\sim
10$. Obviously, the close approximation won't work. How about doing
the following: replace the two black holes at $L/m\sim 10$ by two
black holes at a distance at which a common horizon has just
developed, but with some linear momentum to account to the fact that
they ``fell in'' from $L/m\sim 10$. Evolve the resulting configuration
using the ``close limit''. How do you compute the momentum? They
suggest using Newton theory for two point particles at the equivalent
distance and with the equivalent mass. Evidently, there is a very
crude approximation done: all the energy radiated during the in-fall is
neglected. Yet, as can be seen from the figure, their approximations
works very well for large separations. It becomes identical with the
close approximation when the two black holes start sharing a common
apparent horizon, that explains the strange shape of the
Abrahams--Cook curve. One could expect that this approximation would
be sensitive to the point where one decides to stop the Newtonian
evolution and start to use the ``close limit''. This was explored by
Baker and Li \cite{BaLi} and the result obtained was that the
approximation was quite robust.

\begin{figure}
\centerline{\psfig{figure=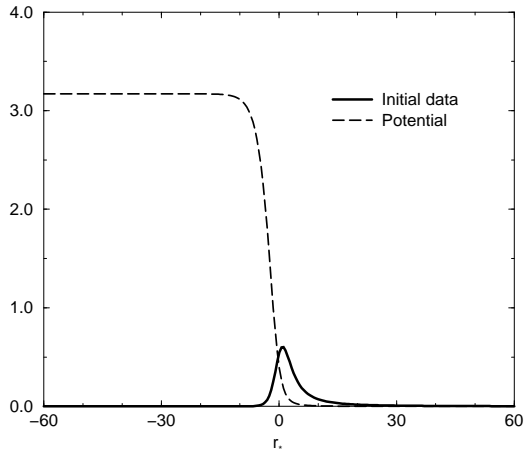,height=70mm}}
\caption{The perturbative initial data for two colliding black holes
and the Zerilli potential. As can be seen, most of the initial data
lies at the left of the potential maximum and will tend to be
swallowed by the black hole. This is the reason why linearized theory
works well in what one would naively presume to be a highly
nonlinear situation.}
\label{fig2}
\end{figure}

The immediate questions seems to be, why does it work well? Wasn't the
collision of two black holes one of the most nonlinear nontrivial
problems in general relativity? How could linearized theory even come
close to predicting it? A possible
answer can be seen in figure \ref{fig2}. Here we show the initial data
and the Zerilli potential plotted as functions of the $r_*$
coordinate. As we see, most of the initial data lies within the
maximum of the Zerilli potential. As such, it will tend to in-fall into
the black hole. Only a small amount escapes to the exterior. Therefore
it is not surprising that linearized theory could do a good job. The
horizon is actually working for us by ``swallowing'' the highly
nonlinear portions of the problem into the inaccessible regions of the
black hole and leaving for the radiative part a weak field, well
described by perturbation theory.

The agreement between numerical and approximate predictions is not
only confined to energies. There is remarkable agreement in waveforms
as well. This is even more surprising, since waveforms have a wealth
of information about the collision and again, there is no parameter to
fit. We will postpone comparisons of waveforms till we discuss the
next section for reasons of space. A long discussion of these
comparisons can also be found in \cite{etaletal}.

So with the close approximation and the Abrahams-Cook extension one
can cover very well (at least qualitatively) the collisions of
momentarily stationary black holes. The question then is, can one make
progress for other types of collisions? In particular, can one say
something about collisions of the physically realistic
``in-spiralling'' type? We will discuss this in the following sections.

\section{Endowing the formalism with error bars}

The first problem at hand is the following: yes, one has an
approximation. But the domain where one can be ``really sure'' the
approximation is going to work is relatively uninteresting (black
holes too close). On the other hand, the results of the test case
suggests the approximation works well somewhat beyond this domain.
Could one somehow get a better grasp of where the domain of validity
lies? The answer we propose to this issue, in collaboration with
Reinaldo Gleiser, Oscar Nicasio and Richard Price is to use second order
perturbation theory. First order perturbation theory has no way to
tell when it is valid. It is only by transcending it and going to
second order that one can get a validity check. More important, one
can make the validity check useful in terms of the physics one is
interested in. For instance, if we are interested in computing
radiated energies and waveforms, then we will demand that the second
order corrections in those quantities be smaller than the first order
ones in order for first order perturbation theory to make sense. The
second order corrections can therefore be seen as ``rough error bars''
of the first order theory predictions. It could happen that other
quantities that we are not interested in (for instance, appropriately
defined invariant quantities close to the horizon) have quite large
second order corrections. But as long as for the quantities of
physical interest one can show that higher order corrections are
smaller than first order theory predictions, then one has reason to
believe those predictions. 

\begin{figure}
\centerline{\psfig{figure=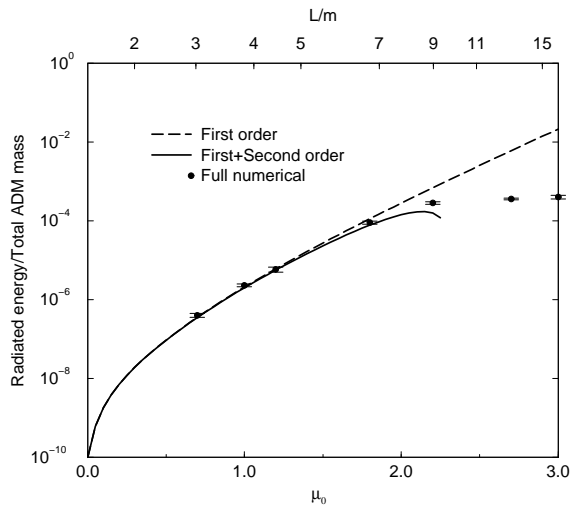,height=70mm}}
\caption{The radiated energy for the head-on collision of momentarily
stationary black holes, computed to first and second order in
perturbation theory compared with the full numerical results. It is
clearly seen that second order theory works very well as an ``error
bar'' of the first order formalism.}
\label{fig3}
\end{figure}

Unfortunately, the formalism of higher order perturbations of black
holes is much less studied than the first order one. Tomita and Tajima
\cite{ToTa} had explored in the seventies second order perturbations,
but they did it in a Newman--Penrose approach for studying fields near
the horizon. We need a $3+1$ point of view in order to supply the
initial data. More importantly, we need a formalism that defines
consistent waveforms far away from the black holes. We were able to
develop such a formalism. A detailed treatment can be found elsewhere
\cite{GlNiPrPucqg,GlNiPrPucmp}. Basically we repeated step by step the
gauge-fixed derivation of the Zerilli equation originally pursued by
Zerilli, keeping up to second order terms. The derivation takes place
in the Regge--Wheeler gauge, which can be fixed to all orders. One
ends up with an equation for a quantity that bears the same
relationship on the second order perturbations as the Zerilli function
did on the first order perturbations. The equation is exactly the same
as the first order Zerilli equation, with the same potential. The only
difference is that the equation involves a ``source term'' that is
quadratic in the first order perturbations and that one can rewrite
entirely in terms of the first order Zerilli function and its spatial
and time (up to third) derivatives. The details can be found in
\cite{GlNiPrPucqg}. However, this is not enough. First of all, the
``second order Zerilli function'' we have just constructed is {\em
not} a ``second order correction'' to the first order Zerilli
function. They are not terms in the expansion of an exact
quantity. They are just combinations of the metric perturbations (in
one case first order, in the other case second order) that carry the
gauge invariant information of the perturbations (this is not manifest
in the gauge fixed approaches we pursue here, but gauge invariant
approaches can be constructed as well, in which these functions are
gauge invariant). So suppose we compute these functions and their time
evolution. What good are they? Without further work, they are of
little use. Since they are not the terms of a series expansion, one
cannot use them to estimate if perturbations are large or small, i.e.,
one cannot compare them with each other. What one can do is try to
compute waveforms in terms of them. What we did for this was to write
the components of the perturbative metric tensor not in the
Regge--Wheeler gauge but in a gauge that is manifestly asymptotically
flat. In such a gauge one can read directly from the metric functions
the ``waveforms'' an observer far away would measure. One can
explicitly compute the relationship between these metric functions to
first and second order and the first and second order Zerilli
functions we computed. One can then see how much second order terms
correct the first order predictions. This is what we were looking
for. As a by-product, using the Landau--Lifshitz pseudo-tensor, it is
immediate to write expressions for the radiated energy in terms of
these asymptotically-flat gauge metric components. This construction
was all explicitly done in \cite{GlNiPrPucqg} and it grossly exceeds
the space we have here to describe it in detail. So we just summarize
it by looking at the results from it. In figure \ref{fig3} we show the
radiated energies for the same problem we discussed in the previous
section, the head-on collision of momentarily stationary black holes
(``the Misner problem'') but now keeping second order terms
\cite{GlNiPrPuprl}.

So we now have a methodology. That is, we can compute the evolution of
a given set of initial data for colliding black holes and use second
order perturbation theory to estimate how good an approximation to the
real result we get. We are therefore ready to attack collisions of
further physical interest. 

Before we move on to non-momentarily stationary cases, let us mention
for completeness that other forms of stationary collisions have also
been treated with the close approximation. Abrahams and Price
\cite{AbPrbl} considered the use of ``Brill-Lindquist'' \cite{BL} type initial
data. These families of initial data are very similar to Misner's, but
the boundary condition given at the throats is different. In the
Misner solution, one requires the geometry to be isometric through the
two holes (as if they were connected by a single ``handle'') whereas
in the Brill-Lindquist case one is considering two holes with
different asymptotically flat interior regions. The two boundary
conditions, when boiled down to the initial value for the Zerilli
equation, only differ in a constant number characterizing the
quadrupole term in the expansion in $1/r$ of the conformal
factor. This difference becomes more important the closer the black
holes are. In spite of being ``in the close limit'', for separations
of $\mu_0 >0.8$ or so, the two problems differ in less than a few
percent. Andrade and Price \cite{AnPr} considered also the collision
of two Brill--Lindquist black holes of unequal mass in the close
limit. This allowed them to probe questions such as ``is there an
optimal mass relation that maximizes radiation?''. Very recent
full numerical work of Anninos and Brandt \cite{AB} apparently
confirm very well the findings of Andrade and Price. Lousto and Price
\cite{LoPr} considered the extreme case of one of the black holes
being infinitesimally small. This makes contact with previous studies
\cite{DRIPP} of particles in-falling into black holes. These studies
help elucidate certain questions about the initial data. For instance,
what sort of boundary conditions for the initial data of two black
holes leads in the particle limit to inadequate answers, etc. The
conclusion seems to be that, at least for the point-particle limit,
the Bowen--York initial data do not work very well. In view of this,
Lousto and Price \cite{LoPr} have proposed enhanced initial data that
is worthwhile exploring further. All of the above studies have been
performed using first order perturbation theory only. Using second
order perturbation theory, Gleiser \cite{Gl} was able to show that the
radiated energy and the change in the ADM mass of the system coincide
in perturbation theory. These kinds of investigations open the road
towards trying to address the back reaction problem in the context of
perturbation theory, and more results are expected in the future.

\section{Non-momentarily-stationary head-on collisions}

The first problem one encounters when trying to address the issue of
non-momentarily-stationary collisions is that there are no known exact
solutions to the initial value problem of general relativity involving
non-vanishing extrinsic curvature. One therefore resorts to the
traditional conformal approach. There, one assumes that the
three-metric is conformally flat $g_{ab} = \phi^4 \delta_{ab}$, and to
simplify things we take maximal slicing ${\rm Tr}K =0$. If one defines
a conformally related extrinsic curvature $\hat{K}_{ab} = \phi^{-2}
K_{ab}$ then the diffeomorphism and Hamiltonian constraints simplify
to,
\begin{eqnarray}
\nabla_a \hat{K}^{ab} &=& 0\label{dif}\\
\nabla^2 \phi &=& -{1\over 8} {\hat{K}^{ab} \hat{K}_{ab} \over \phi^{7}}
\label{Hamil}
\end{eqnarray}
where all the derivatives are with respect to flat space.

The usual strategy to solve these equations has been as follows (see
for example, the work of Bowen and York \cite{BoYo}): construct a
solution to (\ref{dif}) for a single hole in the situation of
interest. For instance, Bowen and York present extrinsic curvatures
that represent situations that have non-vanishing ADM linear momentum
and/or angular momentum. Once you have the extrinsic curvature for a
single hole, just superpose two of them to get the extrinsic curvature
for two black holes. The momentum constraint is linear, so the
resulting extrinsic curvature is still a solution to (\ref{dif}). Plug
the resulting extrinsic curvature in (\ref{Hamil}) and solve the
nonlinear elliptic PDE for $\phi$ (usually numerically). Abrahams and
Price \cite{AbPrfo} have studied the issues associated with taking
numerical initial data and evolving it using perturbation theory. In
fact, the Abrahams--Cook results we discussed before were actually
obtained this way.

Here we will pursue a separate avenue \cite{Pu}. We will solve the
constraint equations {\em approximately}. After all, our whole
formalism is approximate, so we do not really need exact initial
data. We will consider the ``slow'' approximation, where the extrinsic
curvature is small. In such a case, one can ignore the right-hand-side
of the Hamiltonian constraint (\ref{Hamil}) and immediately construct
a solution to the initial value problem: the extrinsic curvature is
the one given by Bowen and York and the conformal factor one takes as
if {\em there were no extrinsic curvature}. That is, we can borrow the
solution of Misner or Brill--Lindquist. From there on, we just apply
the techniques we described in the previous sections to do the
evolution. If one wants to go to second order, the solution we just
presented is a bit too crude, but one can refine it by ``feeding
back'' the extrinsic curvature in the right-hand-side of the
Hamiltonian constraint and solving for the conformal factor to second
order in the momentum only. This can usually be pursued (at least at
each multipolar order $\ell$) analytically.

\begin{figure}
\centerline{\psfig{figure=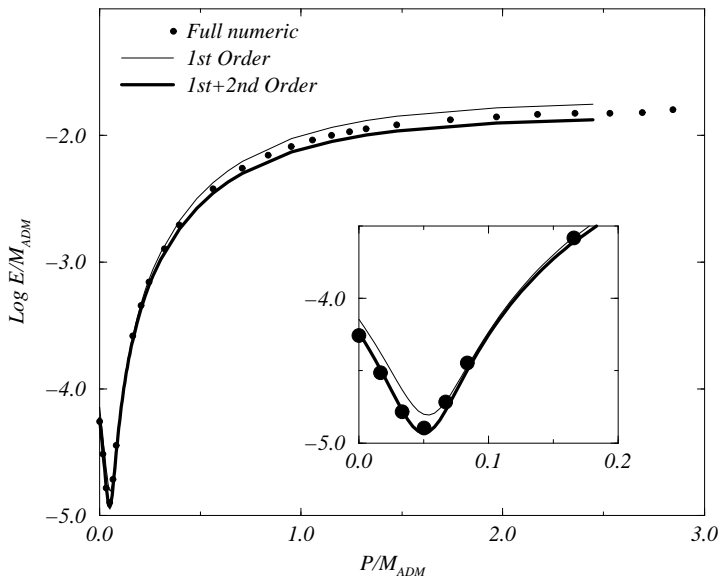,height=90mm}}
\caption{The radiated energy by the head-on collision of two black
holes boosted towards each other, as a function of the momentum, for a
separation of $\mu_0=1.5, L/m=4.2$. We see good agreement with
numerical results for a large range of values of the momentum. We also
see the ``dip'' that occurs for small values of the momentum.}
\label{figboost1}
\end{figure}

Can this possibly work? The best answer is given by seeing some
results \cite{boost1}. Let us take a look at the radiated energies for
the head-on collision of two black holes boosted towards each other
with linear momentum $P$. Figure \ref{figboost1} shows the energy
for a given initial separation, in terms of the Misner parameter
$\mu_0=1.5$. For momentarily-stationary solutions, this was a
respectable separation, but not too large (the black holes are
surrounded by a common event horizon, not by a common apparent
horizon).

In the figure we see that the energies agree very well with the
numerical simulations. But contrary to expected, the approximation
does not deteriorate rapidly for large values of $P$. How could this
be? Weren't we performing a ``slow'' approximation? Evidently there is
a surprise here. The surprise is that indeed, our expression for the
conformal factor should only work well for small values of $P$. But in
the initial data, if one increases $P$, the extrinsic curvature grows
linearly in $P$, whereas the conformal factor ---tied up by the
nonlinearities of the Hamiltonian constraint--- only grows very slowly
with the momentum. As a result, for large values of $P$, the initial
data become ``momentum dominated'' i.e. swamped by the extrinsic
curvature. Since we are inputting the extrinsic curvature {\em
exactly}, then our approximation actually becomes {\em somewhat
better} for large values of the momentum. The reason this is not
entirely true (and we actually see in the plot that the discrepancy
increases with $P$) is that we are inputting exactly $\hat{K}$, the
physical extrinsic curvature inherits some error through the conformal
factor. But if one is not interested in a great degree of accuracy, it
is remarkably simple to provide initial data for this problem. One can
set the conformal factor to unity and get reasonable results! 

This ``momentum dominance'' may be an observation that will lead to a
lot of implications for realistic collisions. After all, one does not
expect that the final moments of a realistic black hole collisions
will be well approximated by slices that resemble the Bowen and York
initial data (which were constructed in an ad-hoc way). Therefore,
there were a lot of worries about what sort of initial data to
actually use in numerical simulations, at least if the latter were not
to start from very far apart ---as surely the initial ones will
not.--- However, momentum dominance may imply that the situation is
not so serious, since for black holes with large values of the
momentum ---as one expects in realistic collisions---, the ambiguities
in the conformal factor of the initial data, could be completely
irrelevant, at least if one is interested in performing computations
that are not too accurate.

The plot for the energies also shows the ``dip'' structure for small
values of the momentum. This is very counterintuitive. One boosts two
black holes towards each other and when they collide they radiate less
(in fact almost an order of magnitude less) than if one had let them
in-fall from rest. This is a peculiarity of the Bowen--York family of
initial data. When one starts boosting the black holes towards each
other, one adds extrinsic curvature to the initial
data. Waveform-wise, the extrinsic curvature happens to be $180^\circ$
out of phase with the conformal factor. Therefore at the beginning
they tend to cancel each other. Past a certain point, the extrinsic
curvature dominates and the energy increases again. Physically, it
appears that a competition is taking place between the added
distortions to the geometry the additional momentum supplies and the
added size of the horizon which is observed in these families of
initial data for increasing momentum \cite{AbComo}. It is like the
horizon ``jumps out'' faster and engulfs more spacetime that could be
radiated when one increases the momentum and decreases the amount of
radiation produced. It is worthwhile noticing that this effect is
linear in $P$, that is, if one reverses the momentum boosting the
black holes apart from each other, it is not observed. 

Due to reasons of space we can again not detail the second order
treatment of boosted black holes we have carried out with Reinaldo
Gleiser, Oscar Nicasio and Richard Price. Details can be seen in
\cite{GlNiPrPuboost2}. Let me just mention that there are several
technical complications in the boosted case. It took us almost two
years to get the formalism in place. To begin with, the initial data
has non-vanishing $\ell=0$ components. One can ignore these to first
order ---they are pure gauge---, but to second order one needs to take
into account the quadratic terms they give rise to. In order to use
the formalism we had (which did not include them), one needs to gauge
away the $\ell=0$ first order term keeping second order terms in place
\footnote{A first order gauge transformation produces higher order
changes in the metric components that are discarded if one works to
first order only, here we need to keep them. Bruni et al have recently
been studying higher order gauge transformations \cite{Bruni}.}. This
is expensive computationally, but eventually was done.  Another issue
that arose is that when one considers $\ell=0$ terms, the
Regge--Wheeler gauge is not unique. We had to perform a further gauge
fixing that is tantamount to a ``variable time shift'' in the
resulting waveforms. Details can be seen in
\cite{GlNiPrPuboost2}. Moreover, the whole issue of doing perturbation
theory when one has more than one parameter (in our case, initial
separation and momentum) in the problem, presents additional
challenges and can lead to inconsistencies if one is not careful.

\begin{figure}
\centerline{\psfig{figure=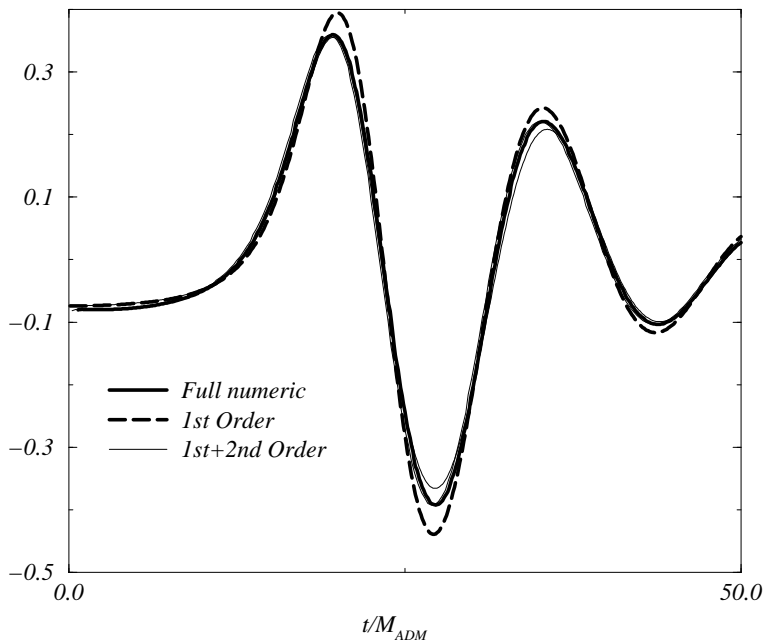,height=100mm}}
\caption{Radiated waveforms for boosted black holes. We see excellent
agreement between the predictions of first and second order
perturbation theory and the full numerical results, in spite of the
fact that this is a very disfavorable case for perturbation
theory. The momentum is very large, $P/M_{ADM}\sim 2.44$.}
\label{figboost3a}
\end{figure}

The end result, however, is very encouraging. Due to reasons of space
I will show only one set of waveforms, for large values of the
momentum. Here is where our approximation works the worst, and yet, as
can be seen in figure \ref{figboost3a}, the agreement with the full
numerical simulations is remarkably good. In fact, it is probably
inconsistent to request any more accuracy from perturbation theory for
so large a value of the perturbative parameter.

What about black holes with spin? One can consider head-on collisions
of spinning black holes rather straightforwardly. For reasons of time,
we have not yet completed the analysis of this problem. In fact, this
problem appears as much simpler than the boosted black holes problem:
there are no $\ell=0$ components, there is no ``mixing'' between the
extrinsic curvature and the conformal factor (at least in first
order), since they are of different parity, so there is no ``dip'' in
the energy. Up to now we have only explored it in first order
perturbation theory. Results can be seen in \cite{HP}. However,
because of normalization and other issues, these results are only
trustworthy for very small values of the momentum at the moment. In
the forthcoming months we expect to present a full analysis including
second order results. An interesting development is that Brandt
\cite{Br} has a full numerical code (developed with Seidel) for
evolving distortions of a Kerr black hole and we are working closely
with him to compare our evolutions with his for similar families of
initial data representing collisions of counter or co-rotating
spinning black holes. A result already visible in \cite{HP} that might
be of interest is that if one aligns the spin along the collision
line, the emitted radiation is the maximum. This contradicts a usual
expectation, i.e., that axisymmetric situations radiate less than
fully non-symmetric ones. This might be an interesting issue to probe
as full 3D codes become more stable, since it is an ``almost 2D''
three dimensional issue.

\section{The problem of initial data}

\begin{figure}
\centerline{\psfig{figure=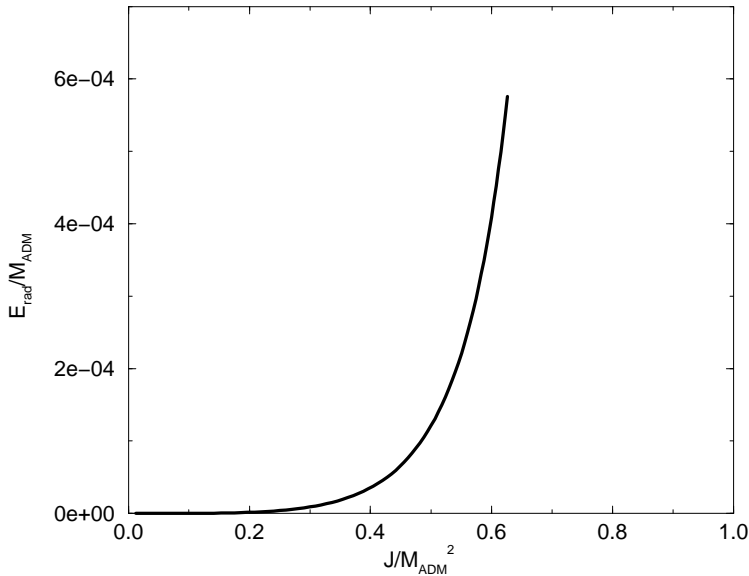,height=90mm}}
\caption{The energy radiated by a {\em single} spinning Bowen--York
black hole as it ``relaxes'' to the Kerr spacetime, as a function of
the spin of the hole. We see that for spins greater than $0.6$, it can
be comparable to the total energy radiated in a head-on collision.}
\label{fig5}
\end{figure}

As has transpired in the previous sections, the initial data that one
constructs to consider black hole collisions following the Bowen--York
method might not be what one needs to represent realistic
situations. Yet, these data are widely used. Therefore, it is good to
get a firm grasp on their physical properties. In particular, it is
known that if one considers a {\em single} boosted or spinning
Bowen--York black hole, one does not end up with a slice of either
boosted Schwarzschild nor Kerr. That is, these families of initial
data represent boosted/spinning black holes ``plus gravitational
radiation''. How much radiation is an issue that becomes more pressing
the closer one needs to start the collision. For the first numerical
simulations it will probably be important, and it is definitely of
interest in the close limit. We therefore set out with Reinaldo
Gleiser, Oscar Nicasio and Richard Price to estimate the amount of
initial radiation present in these solutions to the initial value
problem. To do it, we considered a single spinning Bowen--York black
hole, and evolved it treating it as a perturbation of Schwarzschild
\cite{GlNiPrPuboyo}. As long as the angular momentum is small, this is
a valid approximation. First order perturbation theory does not yield
anything since the first order perturbations are stationary (they
account for the fact that one is dealing with a Kerr rather than a
Schwarzschild black hole). Nontrivial radiation appears only at second
order. With our formalism, we can readily compute this. The result is
summarized in figure \ref{fig5}.
We see that for spins bigger than $0.6$ in units of ADM mass squared,
the radiation present per hole in the initial data can be as large as
that emitted in a collision. Unfortunately our technique is not
applicable for large values of the spin, but still the result is a
cautionary note concerning the use of Bowen--York initial data for
large values of momenta.

\section{In-spiralling collisions}

What can be said about in-spiralling collisions? If one keeps the
total angular momentum small, one can even treat these types of
collisions as perturbations of a Schwarzschild black hole. One could,
for instance, consider the initial data for non-head-on boosted black
holes of Bowen--York type and evolve it as a perturbation of
Schwarzschild. This was done for first order perturbations in
\cite{HP}. These results are only credible for quite small angular
momenta until we have performed a second order analysis. One sees a
similar behavior as in the case of spinning black holes, but the
limited range of momenta involved do not allow to draw many
conclusions or behaviors. In particular, one really expects more
radiation in the case of in-spiralling collisions.

What about trying to treat the problem as a perturbation of the Kerr
spacetime? Would this allow to consider cases with larger angular
momentum? Perturbations of Kerr have been importantly studied, and the
Teukolsky equation is the formalism one would try to use. However,
there are several practical and conceptual difficulties that get in
the way of trying to do this.

The first practical difficulty is that there was not much experience
in working with the Teukolsky equation in the time domain. In such a
domain the equation is not entirely separable, and therefore one is
left, even after decomposing in spherical harmonics, with a difficult
hyperbolic $2+1$-dimensional partial differential
equation. Fortunately, work by Krivan, Laguna, Papadopoulos and
Andersson \cite{KLP}, who wrote a code for integrating the resulting
PDE, now allows us to have at our disposal a tool for the evolution of
perturbations of a Kerr black hole that makes the task almost as easy
as with the Zerilli equation.

The next  two difficulties have to do with properties of the
Bowen--York initial data. The first property is that if one considers
the extrinsic curvatures of two black holes boosted towards each other
in a non-head-on fashion, one can see that it can be rewritten as the
extrinsic curvature of a single spinning black hole of angular
momentum $L=P\times D$ with $P$ the momenta of each hole and $D$ the impact
parameter, plus ``perturbations'' \cite{Yofr}. Prima facie, this is good. It
appears to behave as one expected. However, one quickly notices that
both the extrinsic curvature of the spinning black hole and ``the
perturbations''  are of order $P\times D$. That is, if one was hoping
that using perturbations of a Kerr hole would allow to treat higher
angular momentum cases by ``absorbing'' the extra angular momentum
into the background, this shows that it is not the case. If one
increases the angular momentum, perturbations become large. This
strikes a blow to the hope that one could do much better using
perturbations of Kerr instead of simply perturbations of
Schwarzschild. It might be the case that one can do better, but it
will not be spectacular. Given the extra complications implied by the
Kerr case, it even raises the question of why investing the extra
effort for what will appear to be a small payoff. Of course, we can
never be sure of how much better we can do until we try it. It could
be that numerical coefficients conspire to be small in the right
direction, in spite of the terms being of the same ``order'' in terms
of the angular momentum. We shall see. 

The other difficulty associated with the Bowen--York type of initial
data is related to the fact that the three-metric is conformally
flat. There is no explicitly known slicing of the Kerr metric with
conformally flat slices. Worse, the Teukolsky equation is explicitly
derived in Boyer--Lindquist coordinates, where the slicing is not
conformally flat. How does one use the Bowen--York initial data in
this formalism? One possibility is to just ignore the issue and define
as a perturbation the departure from Boyer--Lindquist that makes the
metric conformally flat. This could lead to artificially large
perturbations, since one is treating as a perturbation a mismatch in
slicing. Again, for small values of the angular momentum, the spatial
part of the Kerr metric is approximately conformally flat and
therefore this should not be a bad approximation. Treating the
Bowen--York data in this way is currently being pursued by Baker,
Khanna, Laguna and Pullin \cite{BaKhLaPu}.

Another way around this latter issue would be to use {\em different}
families of initial data from that of Bowen and York. It turns out
that this is not that easy to do. If one relinquishes conformal
flatness, all constraints become coupled and it is difficult to make
any sense of a ``superposition principle'' that would allow to set
initial data for two black holes. Some steps in this direction have
been recently taken by Baker and Puzio \cite{BaPu} and independently
by Krivan and Price \cite{KrPr}. Their constructions at the moment
only work for axisymmetric data. But they are non-conformally flat,
and have the attractive property of yielding Kerr black holes for far
away black holes and a single Kerr black hole in the close limit. At
least for head-on collisions of spinning black holes, these new
families of initial data hold a lot of promise. Evolution of them is
now being studied by the above four authors using the Teukolsky
numerical code mentioned above.

Finally, a more technical, but yet practically important issue associated
with perturbations of Kerr is that because they had never been studied
in the time domain, no one had derived formulas giving the Teukolsky
function and its time derivatives in terms of a three metric and an
extrinsic curvature. The calculation is complicated given the
nontrivial nature of the background, but by now it has been completed
independently by Campanelli, Lousto, Krivan, Baker and Khanna
\cite{CaLoKrBaKh}, who now have formulae suitable for practical use.

Summarizing, we can expect over the next year or so progress in close
limit of in-spiralling collisions. The extent to which one can push
these calculations will depend on how the forthcoming results shape
up. One can envision several internal consistency checks (for instance
treat the problem as a perturbation of Schwarzschild, to first and
second order, compare with the same problem treated as a perturbation
of Kerr), and depending on the results we will see how useful the
close approximation will be for these, the most physically realistic
cases. 

\section{Conclusions}

We have attempted to make the case that studying black hole collisions
in the close limit using perturbation theory could be a valuable tool
for understanding the physics involved. In spite of the fact that it
will probably never give the complete picture of a collision, the use
of the close approximation {\em opens new insights} that were normally
unforeseen in the complete absence of a technique to compute the
collision. In fact, certain aspects may even remain obscure after a
full numerical simulation. The kind of handle one gets on issues when
one can treat a problem approximately normally complements the
knowledge one gains through full numerical simulations. 

A completely uncharted area that warrants further exploration is the
use of these techniques in the realm of neutron stars (or neutron
star-black hole) collisions. There are evident possibilities of
applying the close limit idea and trying to gain insights into these most
complex collisions. This is a major effort that we do not expect to
get fully into until further advancing the binary black hole
program. But it is an evident next step in the future.

Maybe the over-encompassing lesson learnt from this approach is that
the problem of binary collisions of compact objects is hard enough for
it to benefit from the synergy of various approaches, in particular,
full numerical simulations and approximate calculations. In the case
of binary black holes we see that taking place right now.

\section*{Acknowledgements}

This talk described work carried along primarily with Reinaldo
Gleiser, Oscar Nicasio and Richard Price. Part of the work was also
completed with John Baker and Hans-Peter Nollert.  The numerical synergy
would not have been possible without the continuous support of the
NCSA-Potsdam-WashU group, especially Peter Anninos, Steve Brandt, Ed
Seidel and Wai-Mo Suen. I also wish to thank Reinaldo Gleiser, Carlos
Lousto and Hans-Peter Nollert for comments on this manuscript. The
work of JP is supported in part by grants NSF-INT-9512894,
NSF-PHY-9423950, NSF-PHY-9507686, the Pennsylvania State University,
the Eberly Family Research Fund at Penn State and the Alfred P. Sloan
foundation.

\end{document}